\documentstyle[amssymb,graphicx]{article}

\begin{document}
\title{Sensitivity to initial conditions at bifurcations in one-dimensional
nonlinear maps: rigorous nonextensive solutions}
\author{F. Baldovin$^{\dagger }$ and A. Robledo$^{\ddagger }$ 
\thanks{E-mail addresses: baldovin@cbpf.br, robledo@fisica.unam.mx}\\
\it{$^{\dagger }$Centro Brasileiro de Pesquisas F\'{i}sicas}\\
\it{Rua Xavier Sigaud 150,22290-180 Rio de Janeiro -- RJ, Brazil}\\
\it{$^{\ddagger }$Instituto de F\'{i}sica, Universidad Nacional 
Aut\'onoma de M\'exico}\\
\it{Apartado Postal 20-364, M\'{e}xico 01000 D.F., Mexico.} }
\date{September 10, 2002}
\maketitle

\begin{abstract}
Using the Feigenbaum renormalization group (RG) transformation we work out
exactly the dynamics and the sensitivity to initial conditions for unimodal
maps of nonlinearity $\zeta >1$ at both their pitchfork and tangent
bifurcations. These functions have the form of $q$-exponentials as proposed
in Tsallis' generalization of statistical mechanics. We determine the 
$q$-indices that characterize these universality classes and perform for the
first time the calculation of the $q$-generalized Lyapunov coefficient $\lambda
_{q} $. The pitchfork and the left-hand side of the tangent bifurcations
display weak insensitivity to initial conditions, while the right-hand side
of the tangent bifurcations presents a `super-strong' (faster than
exponential) sensitivity to initial conditions. We corroborate our
analytical results with {\em a priori} numerical calculations.
\end{abstract}


The nonextensive generalization \cite{tsallis_01} of the canonical
statistical mechanics has raised interest in testing its applicability in
several suggested circumstances in a variety of physical systems \cite
{tsallis_01}. A class of problems where this issue has been much studied
recently is the dynamical behavior of nonlinear iterated maps under critical
conditions \cite{tsallis_02,costa_01,lyra_01,moura_01,baldovin_01}. Such is
the case of the bifurcation points associated to deterministic chaos in
simple nonlinear dissipative maps, like those occurring in the logistic map
and its generalization to nonlinearity $\zeta >1$. These types of critical
states provide an exceptional opportunity to examine explicitly the
underlying mathematical structure and physical implications of their
scale-invariant, power-law properties. As it is well-known the
period-doubling and intermittency routes to chaos are based on the pitchfork
and tangent bifurcations, respectively \cite{schuster_01}. The pitchfork
bifurcations are the mechanism for the successive doubling of periods of
stable orbits, and their accumulation point, an orbit of infinite period, is the
so-called chaos threshold. The tangent bifurcation is a different mechanism
linking periodic orbits with chaos in which intermittency is a precursor to
periodic behavior \cite{schuster_01}. Like in conventional equilibrium
second order transitions, the bifurcation points have universal properties
that can be obtained by means of renormalization group (RG) techniques,
which for maps of the logistic type are based on functional composition and
rescaling. As a result of this the {\em static} universal properties such as
those derived from fixed-point maps and their perturbation equations are
well understood \cite{schuster_01}. However, the {\em dynamical} properties
and their implications are only now being studied in a similar fashion by
the same RG approach \cite{robledo_01}. Here we present RG analytical
results for the universal dynamics at bifurcation points of unimodal
(unidimensional with a single maximum) maps of arbitrary nonlinearity $\zeta
>1$.

\begin{figure}
\begin{center}
\includegraphics[width=8cm,angle=0]{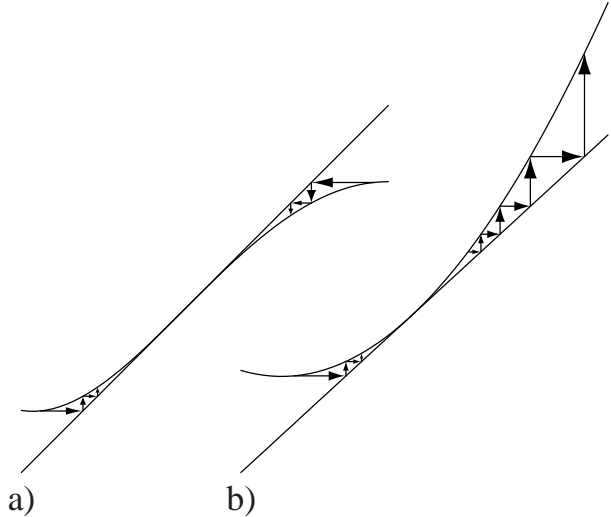}
\end{center}
\caption{\small 
Schematic form of $f^{(n)}$ at the a) pitchfork and b) tangent
bifurcations.}
\label{schematic}
\end{figure}
At each of the map bifurcation points the sensitivity to initial conditions 
$\xi _{t}$ obeys power law behavior for large iteration time $t$ in contrast
to the usual exponential growth or decay generic of noncritical states \cite
{tsallis_02}. This of course implies the vanishing of the Lyapunov $\lambda
_{1}$ coefficient that measures either periodic ($\lambda _{1}<0$) or chaotic
behavior ($\lambda _{1}>0$). The generalized nonextensive theory offers an
attractive alternative to describe the dynamics at such critical points \cite
{tsallis_02}, the sensitivity $\xi _{t}$ would in these cases obey the 
$q$-exponential expression with a $q$-generalized Lyapunov coefficient 
$\lambda _{q}$
\begin{equation}
\xi _{t}=\exp _{q}(\lambda _{q}t)\equiv [1-(q-1)\lambda _{q}t]^{-
\frac{1}{q-1}}\;\;\;\;(q\in \mathbb{R}),  
\label{sensitivity_00}
\end{equation}
that yields the customary exponential $\xi _{t}$ with $\lambda _{1}$ when 
$q\rightarrow 1$. 
The definition of $\lambda_q$ differs from other generalizations of a Lyapunov
coefficient \cite{beck_01}.
One of us has recently pointed out \cite{robledo_01} a
clear connection between the fixed-point RG solutions at bifurcation points
of unimodal maps and nonextensive entropy extremal properties. The aim of
this letter is to work out analytical solutions for the dynamics and the
sensitivity to initial conditions in the proximity of the first bifurcations
of both types, and to corroborate these results numerically. In doing this
we confirm previous estimations of the nonextensive parameter $q$ and
calculate for the first time the $q$-generalized Lyapunov coefficient $\lambda _{q}$
in terms of the map parameters.

Let us start by briefly recalling the background formalism. We consider the
generalization to nonlinearity of order $\zeta >1$ of the logistic map 
\begin{equation}
f_{\mu }(x)=1-\mu |x|^{\zeta },  \label{original_map}
\end{equation}
where $-1\leq x\leq 1$ and $0\leq \mu \leq 2$. These maps display a
``pitchfork bifurcation regime'' for $\mu <\mu _{\infty }$ where $\mu
_{\infty }$ is the value of the period-doubling accumulation point (onset of
chaos). On the other hand, as a consequence of the tangent bifurcations the
''chaotic regime'' for $\mu >\mu _{\infty }$ becomes interrupted at certain
values of $\mu $ by windows of periodic behavior \cite{schuster_01}. The
solution of Hu and Rudnick \cite{schuster_01,hu_01} to the Feigenbaum RG
recursion relation \cite{feigenbaum_01} 
for the tangent bifurcation was obtained as follows. For
the transition to a periodic window of order $n$ there are $n$ points for
which the original map $f$ is tangent to the line of unit slope. Choosing
one of these points, shifting the origin of coordinates to this point, and
making an expansion of the $n$-th composition of the original map, one
obtains 
\begin{equation}
f^{(n)}(x)=x+u|x|^{z}+o(|x|^{z}),  \label{expansion}
\end{equation}
where $u$ is the leading expansion coefficient. The RG fixed-point map 
$x^{\prime }=f^{*}(x)$ was found to be 
\begin{equation}
x^{\prime }=x[1-(z-1)u\;{\rm sgn}(x)|x|^{z-1}]^{-\frac{1}{z-1}}.  \label{RG_solution}
\end{equation}
This solution has a power-series expansion in $x$ that coincides with 
Eq. (\ref{expansion}) in the two lowest-order terms. In Ref. \cite{robledo_01} it
was observed that the previous scheme is also applicable to the pitchfork
bifurcations $\left. df^{(2^{k-1})}(x)/dx\right| _{x=0}=-1$ of order 
$n=2^{k} $, $k=1,2,...$, provided that the sign of $u$ is changed for $x>0$.
As it will become clear below, we note that the power $z$ appearing in 
Eqs. (\ref{expansion}) and (\ref{RG_solution}) is different from the power $\zeta $
in the original map Eq. (\ref{original_map}).

Now, taking into account that the fixed-point map Eq. (\ref{RG_solution})
satisfies $f^{*}(f^{*}(x))=\alpha ^{-1}f^{*}(\alpha x)$ with $\alpha
=2^{1/(z-1)}$, we obtain by repeated functional composition the following
remarkable property 
\begin{equation}
f^{*^{(m)}}(x)=\frac{1}{m^{\frac{1}{z-1}}}f^{*}
(m^{\frac{1}{z-1}}x),\;\;\;\;m=1,2,...  \label{RG_relation}
\end{equation}
This property implies that, for a total number of iterations $t=mn$, 
$m=1,2,...$, with sufficiently small initial $x_{0}$, and for $n$ and
therefore $t$ large, the fixed-point map can be written as 
\begin{equation}
x_{t}\equiv \left[ f^{(n)}\right]
^{(m)}(x_{0})=x_{0}[1-(z-1)a\;{\rm sgn}(x_0)|x_{0}|^{z-1}t]^{-\frac{1}{z-1}},  
\label{dynamics}
\end{equation}
where $a\equiv u/n$. This is equivalent to a continuous time $t$
approximation of Eq. (\ref{RG_solution}) obtained by taking $x_{t}=x^{\prime
}$, $x_{0}=x$ and $at=u$. We notice also that by introducing an additional
index $l=1,2,...,n-1$, the same expression holds for 
$x_{t+l}=f^{(mn+l)}(x_{0})$, since $x_{0}$ on the right-hand side of this
equation can be replaced by $f^{(l)}(x_{0})$. In words, the time evolution
of the original map in the neighborhood of a bifurcation point of order $n$,
consists of a bundle of $n$ orbits or trajectories of the form (\ref
{dynamics}). But if one considers instead the time evolution of the 
$n$-composed map $f^{(n)}$ this is given by a single orbit of the form (\ref
{dynamics}), only now with $t=m$ and $a=u$. It can also be verified that Eq.
(\ref{dynamics}) satisfies the property $dx_{t}/dx_{0}=(x_{t}/x_{0})^{z}$,
and this implies that the sensitivity to initial conditions $\xi _{t}\equiv
\lim_{\Delta x_{0}\to 0}(\Delta x_{t}/\Delta x_{0})$ has the form 
\begin{equation}
\xi _{t}(x_{0})=[1-(z-1)a\;{\rm sgn}(x_0)|x_{0}|^{z-1}t]^{-\frac{z}{z-1}}.
\label{sensitivity_01}
\end{equation}
Comparison of Eq. (\ref{sensitivity_01}) with Eq. (\ref{sensitivity_00})
yields the identifications \cite{buiatti_01} 
\begin{equation}
q=2-\frac{1}{z}\;\;{\rm and}\;\;\lambda _{q}(x_{0})=za\;{\rm sgn}(x_0)|x_{0}|^{z-1}.
\end{equation}
Further, by using the known \cite{gaspard_01} form $\rho (x)\sim \left|
x\right|^{-(z-1)}$ for the invariant distribution of $f^{(n)}$ in Eq. (\ref
{expansion}) , we have that the average $\bar{\lambda}_{q}$ of $\lambda
_{q}(x_{0})$ over $x_{0}$ yields the piece-wise constant 
\begin{equation}
\bar{\lambda}_{q}=za\;{\rm sgn}(x_{0}).
\end{equation}
It is interesting to notice \cite{robledo_01} that this average corresponds
to the $q$-extension of the customary expression for the Lyapunov coefficient 
$\lambda _{1}$, obtained as the average of $\ln |df(x)/dx|$ over $\rho (x)$.
Here we have 
\begin{equation}
\bar{\lambda}_{q}=\int dx\rho (x)\ln _{q}\left| \frac{df^{(n)}(x)}{dx}
\right| ,
\end{equation}
where $\ln _{q}(y)\equiv (y^{1-q}-1)/(1-q)$, and $\ln _{q}\left( \exp
_{q}(x)\right) =x$. Likewise, for the $n$-composed map $f^{(n)}$ the
sensitivity to initial conditions is characterized by the $q$-generalized
Lyapunov coefficient $\overline{\Lambda }_{q}=n\bar{\lambda}_{q}=zu$. In Ref. 
\cite{robledo_01} the alternative expression for the sensitivity to initial
conditions was proposed 
\begin{equation}
\xi _{t}=\left[ \exp _{q}(\lambda _{q}t)\right] ^{q}\equiv [1-(q-1)\lambda
_{q}t]^{-\frac{q}{q-1}},
\end{equation}
as suggested by a noteworthy similarity between the RG perturbation
expression for $x_{t}$ and the corresponding nonextensive generalization of
the Kolmogorov-Sinai entropy \cite{robledo_01}. Using this form, the
identifications are $q=z$ and $\bar{\lambda}_{q}=a$.

\begin{figure}
\begin{center}
\includegraphics[width=8cm,angle=0]{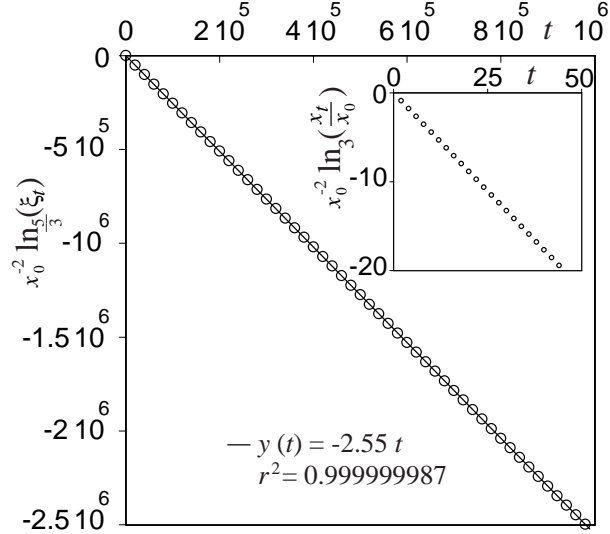}
\end{center}
\caption{
Weak insensitivity to initial conditions for the 1st pitchfork
bifurcation when $\zeta =1.75$. The straight line for the $q$-logarithm of 
$\xi _{t}$ vs $t$ with $q=5/3$ and the slope $m=-2.55$ confirms the
analytical result and the specific values predicted for $q$ and $\lambda
_{q} $ for this transition. The inset shows the corresponding behavior for a
single trajectory.}
\label{1st_pitch}
\end{figure}
Next, we proceed to show explicit results for the index $q$ and the
$q$-generalized Lyapunov coefficient $\bar{\lambda}_{q}$ for the first pitchfork
and tangent bifurcations. 
A pitchfork bifurcation of order $n=2^{k}$ (see Fig. \ref{schematic}a) is characterized by
the following conditions 
\begin{eqnarray}
f^{(2^{k-1})}(y_{c}) &=&y_{c}\;\;\;\;(k=1,2,...), \\
\left. \frac{df^{(2^{k-1})}}{dy}\right| _{y=y_{c}} &=&-1\;\;{\rm and}\;\;\left. 
\frac{df^{(2^{k})}}{dy}\right| _{y=y_{c}}=1,  \nonumber
\end{eqnarray}
these determine the values of the critical parameter $\mu _{c}$ and of the 
$2^{k-1}$ critical positions $y_{c}$. Shift of coordinates ($x\equiv y-y_{c}$
and $x^{\prime }=f^{(2^{k})}(y)-y_{c}$) and expansion yields 
\begin{equation}
f^{(2^{k})}(x)=x+u|x|^{3}+o(x^{3}),
\end{equation}
since $d^{2}f^{(2^{k})}/dx^{2}|_{x=0}=0$ always at these transitions,
independently of the nonlinearity $\zeta $. The coefficient $u$ is then
given by $u=\pm (1/6)d^{3}f^{(2^{k})}/dx^{3}|_{x=0}$, where the $+$ sign
applies for $x>0$ and the $-$ sign for $x<0$, and where 
$d^{3}f^{(2^{k})}/dx^{3}|_{x=0}<0$. As a first result we have that $z=3$ and 
$q=5/3$ for the sensitivity to initial conditions of all the pitchfork
bifurcations of any nonlinearity $\zeta $. Note that for any value of 
$x_{0}\neq 0$ the argument of the $q$-exponential (\ref{sensitivity_01}) is
negative, and this, combined with $q>1$, implies that the pitchfork
bifurcations display {\em weak insensitivity} \cite{tsallis_02} to initial
conditions at both sides of the critical position. In the case of the first
pitchfork bifurcation ($k=1$) its location is given by $\mu _{c}=(\zeta
+1)^{\zeta }/(\zeta ^{\zeta }(\zeta +1))$ and $y_{c}=\zeta /(\zeta +1)$,
whereas the coefficient $u$ is given by 
$u=\mp(\zeta ^{4}+2\zeta ^{3}-2\zeta -1)/(6\zeta ^{2})$.
So that the $\zeta $-dependent $q$-generalized Lyapunov coefficient for $f^{(2)}$
is 
\begin{equation}
\bar{\lambda}_{q}=-\frac{1}{2}\frac{\zeta ^{4}+2\zeta ^{3}-2\zeta -1}{\zeta
^{2}}.  \label{pitchfork_z}
\end{equation}
In Fig. \ref{1st_pitch} we present the numerical results for the first pitchfork
bifurcation with $\zeta =1.75$ that corroborate the analytical prediction 
$\bar{\lambda}_{q}=-2.\,5466$. We have verified Eq. (\ref{pitchfork_z}) for
several values of $\zeta $. 

The location and the expansion coefficient for the second pitchfork
bifurcation ($k=2$) for general $\zeta $ are harder to obtain analytically,
but as an illustration we give results for the logistic map ($\zeta =2$). We
obtain $\mu _{c}=5/4$ and the two critical positions 
$y_{c}=2(1\pm \sqrt{2})/5$. 
Interestingly, $\bar{\lambda}_{q}$ for the first critical position is 
$\bar{\lambda}_{q}=-3(250+125\sqrt{2})/4$, while for the second position is 
$\bar{\lambda}_{q}=-3(250-125\sqrt{2})/4$.

\begin{figure}[tbp]
\begin{center}
\includegraphics[width=14cm,angle=0]{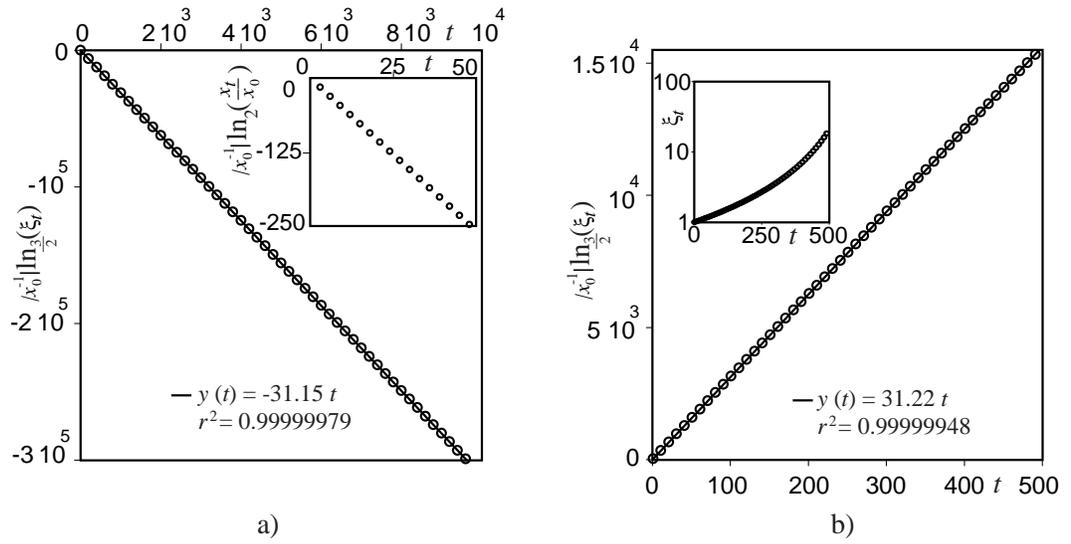}
\end{center}
\caption{\small 
a): Weak insensitivity to initial conditions at the left-hand side of the
1st tangent bifurcation when $\zeta =2$. The straight line for the 
$q$-logarithm of $\xi _{t}$ vs $t$ with $q=3/2$ and the slope $m=-31.15$
confirms the analytical result and the specific values predicted for $q$ and 
$\lambda _{q}$ for this transition. The inset shows the corresponding
behavior for a single trajectory.
b): `Super strong' sensitivity to initial conditions at the right-hand
side of the 1st tangent bifurcation when $\zeta =2$. The straight line for
the $q$-logarithm of $\xi _{t}$ vs $t$ with $q=3/2$ and the slope 
$m=31.22 $ confirms the analytical result and the specific values predicted
for $q$ and $\lambda _{q}$ for this transition. 
The inset shows the logarithm of $\xi_t$ vs $t$ 
where faster than exponential growth can be clearly observed.}
\label{tang}
\end{figure}
The conditions that determine the critical parameter and the $n$ critical
positions of a tangent bifurcation of order $n$ (see Fig. \ref{schematic}b) are 
\begin{equation}
f^{(n)}(y_{c})=y_{c}\;\;{\rm and}\;\;\left. \frac{df^{(n)}}{dy}\right|
_{y=y_{c}}=1,
\end{equation}
so that, by carrying out the same change in coordinates as before, one obtains
the expansion 
\begin{equation}
f^{(n)}(x)=x+u|x|^{2}+o(x^{2}),
\end{equation}
where the coefficient $u$ is now given by 
$u=(1/2)d^{2}f^{(n)}/dx^{2}|_{x=0}>0$. Hence we have $z=2$
and $q=3/2$ for the sensitivity to initial conditions of all the tangent
bifurcations regardless of the value of $\zeta $. The left-hand side of
the tangent bifurcation points exhibits, in analogy to the previous case, a 
{\em weak insensitivity} to initial conditions. However at the right-hand
side of the bifurcation the argument of the $q$-exponential becomes positive
and this, together with $q>1$, results in a {\em `super strong' sensitivity}
to initial conditions, i.e. a sensitivity that is faster than exponential. 
Once again we use the logistic map $\zeta =2$ as an illustration, and we
choose for its first tangent bifurcation ($n=3$) at $\mu _{c}=7/4$ one of
the three critical positions, $y_{c}=0.031405...\;$, and determine its
expansion coefficient to be $u=15.608...\;$. Thus
for $f^{(3)}$ the $q$-generalized Lyapunov coefficient is $\bar{\lambda}_{q}=\pm
31.216...$ for $x_{0}\gtrless 0$. In Fig. \ref{tang}a) we plot the numerical results
for the left side of this bifurcation, while in Fig. \ref{tang}b) we show the `super
strong' sensitivity to initial conditions characteristic of the right side
of the tangent bifurcation. 

Naturally, only for very low order $n$ of the bifurcation and/or special
values of the nonlinearity $\zeta $ it is feasible to obtain algebraic
solutions for the locations of the transitions and then for the leading
expansion coefficient $u$. Nevertheless, it is always possible to implement
numerical methods to determine them, and consequently the $q$-generalized
Lyapunov coefficient $\bar{\lambda}_{q}$. As shown in Figs. \ref{1st_pitch} and 
\ref{tang} the {\em a
priori} numerical calculations for the simplest examples selected here
provide a striking confirmation of our RG predictions for both $q$ and 
$\bar{\lambda}_{q}$.

In summary, we have fully determined the dynamical behavior at the pitchfork
and tangent bifurcations of unimodal maps of arbitrary nonlinearity $\zeta
>1 $. This was accomplished via the consideration of the solution to the
Feigenbaum RG recursion relation for these types of critical points. Our
study has made use of the specific form of the $\zeta $-logistic map but the
results have a universal validity as conveyed by the RG approach. The RG
solutions are exact and have the analytical form of $q$-exponentials, we
have shown that they are the time (iteration number) counterpart of the
static fixed-point map expression found by Hu and Rudnick for the tangent
bifurcations and that is applicable also to the pitchfork bifurcations \cite
{robledo_01}. The $q$-exponential form of the time evolution implies an
analytical validation of the expression for the sensitivity to initial
conditions suggested by the nonextensive statistical mechanics \cite
{tsallis_02}. It also provides straightforward predictions for $q$ and 
$\bar{\lambda}_{q}$ in terms of the map critical point properties 
\cite{robledo_01}. We found that the index $q$ is independent of 
$\zeta $ and takes one of
two possible values according to whether the transition is of the pitchfork
or the tangent type. The $q$-generalized Lyapunov coefficient $\bar{\lambda}_{q}$
is simply identified with the leading expansion coefficient $u$. As we have
shown these predictions are unquestionably corroborated by {\em a priori}
numerical calculations. Significantly, both families of bifurcations display
non-canonical dynamical behavior (either weak insensitivity or `super
strong' sensitivity to initial conditions). The universal attribute of our
results invites some reflection on the rationale for non-canonical dynamics.
This is most apparently linked to the fact that the tangency shape of the
map at these critical points either effectively confines or expels
trajectories causing an abnormal, incomplete, sampling of phase space (here 
$-1\leq x\leq 1$).\smallskip

\section*{Acknowledgments}
We would like to thank C. Tsallis and L.G. Moyano for useful discussions and
comments. AR gratefully acknowledges the hospitality of the Centro
Brasileiro de Pesquisas F\'{\i}sicas where this work was carried out and the
financial support given by the CNPq processo 300894/01-5 (Brazil). AR was
also partially supported by CONACyT grant 34572-E and by DGAPA-UNAM grant
IN110100 (Mexican agencies). FB has benefitted from partial support by
CAPES, PRONEX, CNPq, and FAPERJ (Brazilian agencies).

\end{document}